\documentclass[aps,pre,twocolumn,groupedaddress,showpacs,floatfix]{revtex4}

\usepackage{graphicx}

\begin{document}

\preprint{NSF-KITP-04-51}
\title{Quantum Monte Carlo simulations of confined bosonic atoms in optical lattices}
\author{Stefan Wessel$^{(1)}$} 
\author{Fabien Alet$^{(1,2)}$}
\author{Matthias Troyer$^{(1,2)}$} 
\author{G. George Batrouni$^{(3,4)}$} 
\affiliation{$^{(1)}$Theoretische Physik, ETH Z\"urich, CH-8093 Z\"urich, Switzerland}
\affiliation{$^{(2)}$Computational Laboratory, ETH Z\"urich, CH-8092 Z\"urich, Switzerland} 
\affiliation{$^{(3)}$Institut Non-Lin\'eaire de Nice, Universit\'e de Nice-Sophia Antipolis, France\footnote{Permanent address.}}  
\affiliation{$^{(4)}$Department of Physics, NTNU, N-7491 Trondheim, Norway} 

\date{\today}
\begin{abstract}
We study properties of ultra-cold bosonic atoms in one, two 
and three dimensional optical lattices by
large scale quantum Monte Carlo simulations of the Bose Hubbard model
in parabolic confinement potentials.  Local phase structures of the
atoms are shown to be accessible via a well defined local
compressibility, quantifying a global response of the system to a
local perturbation.  An indicator for the presence of extended
Mott plateaux is shown to stem from the shape of the coherent
component of the momentum distribution function, amenable to
experimental detection. 
Additional fine structures in the momentum
distribution are found to appear unrelated to the local phase
structure, disproving previous claims.  We discuss limitations of
local potential approximations for confined Bose gases, and the absence of quantum
criticality and critical slowing down in parabolic confinement potentials, thus accounting for 
the fast dynamics in establishing phase coherence in current experiments.
In contrast, we find that
flat
confinement potentials allow quantum critical behavior
to be observed already on moderately sized optical lattices. 
Our results furthermore demonstrate, that the experimental detection of the Mott 
transition 
would be significantly eased in flat confinement 
potentials.
\end{abstract}

\pacs{03.75.Hh,03.75.Lm,05.30.Jp}

\maketitle


\section{Introduction}

Experiments on ultra cold atomic gases in optical
lattices~\cite{greiner,esslinger} provide the unique opportunity to
directly compare theoretical studies of strongly correlated many body
quantum lattice models with near-perfect experimental realizations of
those models~\cite{zoller}. Unlike for other strongly correlated
systems, where often the models simulated numerically are prototype
{\it toy models} for describing the properties of such materials, the
same models are here {\it realistic models} of confined cold atoms,
allowing for quantitative comparisons.

So far, experiments have focused on bosonic
atoms~\cite{greiner,esslinger}, which are easier to cool than
fermions. A quantitative understanding of these bosonic systems will,
in the future, allow to correctly interpret measurements on atomic
fermion gases, which could be used as analog quantum simulators for
strongly correlated fermionic systems~\cite{hofstetter}.

Here we present results of an extensive set of quantum Monte Carlo
simulations performed on one dimensional (1D), two dimensional (2D)
and three dimensional (3D) systems of confined bosonic atoms in
optical lattices, extending previous work in 1D~\cite{batrouni} and
3D~\cite{prokofev}. Parts of our results were already presented
elsewhere~\cite{advances}, here we provide a more detailed discussion
and new results on confined Bose gases.  Details on the many body
quantum lattice model used for simulating confined bosons, as well as
the employed quantum Monte Carlo method are discussed in the following
section.

In the first part of the paper, Sec.~\ref{sec:model}, we focus on
simulations of bosons in 2D confinement potentials, and study various
aspects of confined bosons in a realistic setup.  In particular, we
(i) identify a local probe for the state at a given location inside
the inhomogeneous trap, (ii) present a detailed analysis concerning
the limitations of local potential approximations, and (iii) discuss the
nature of spatial correlations inside the trap.

\begin{figure}
\begin{center}
\includegraphics[width=4.2cm]{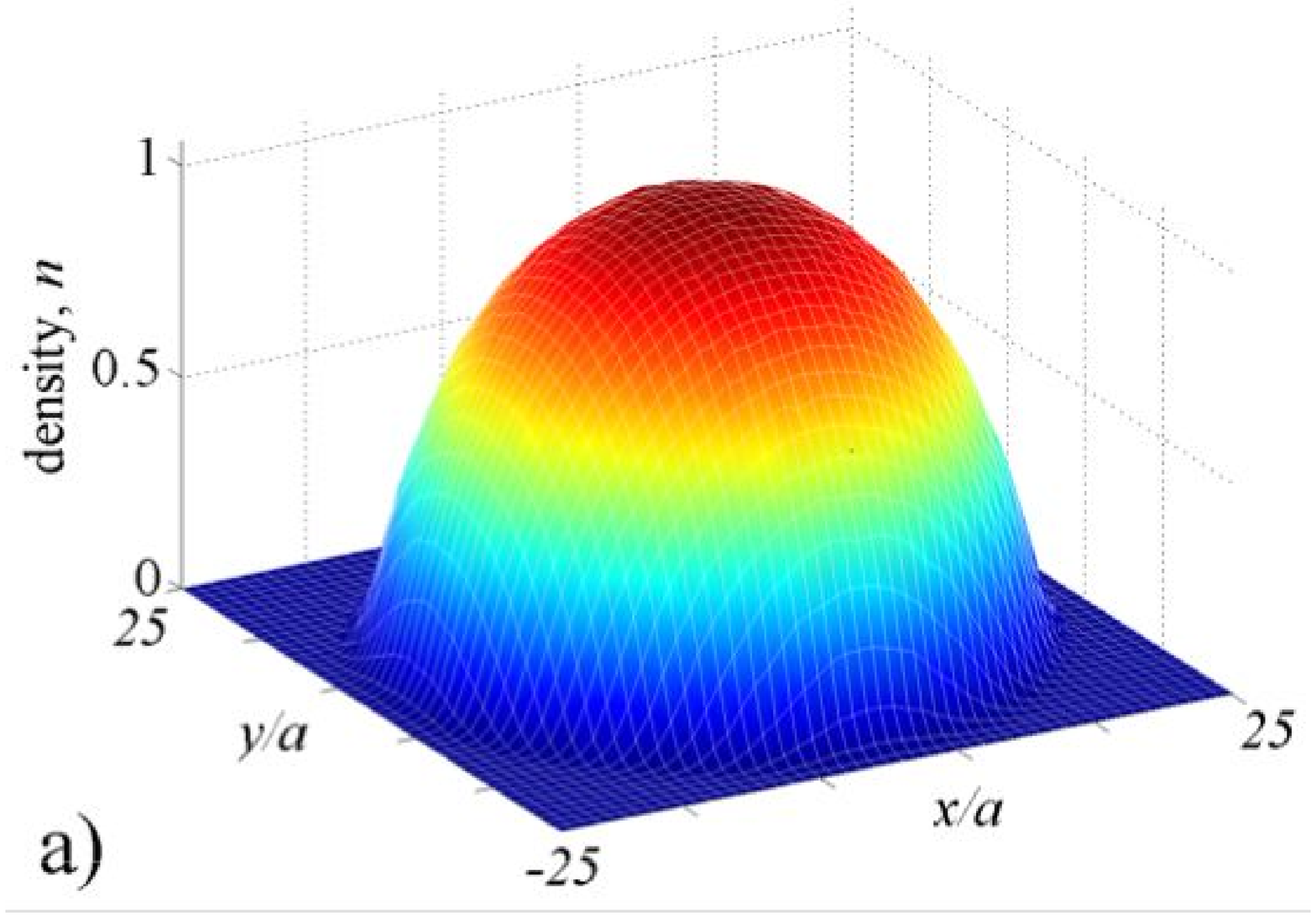}
\includegraphics[width=4.2cm]{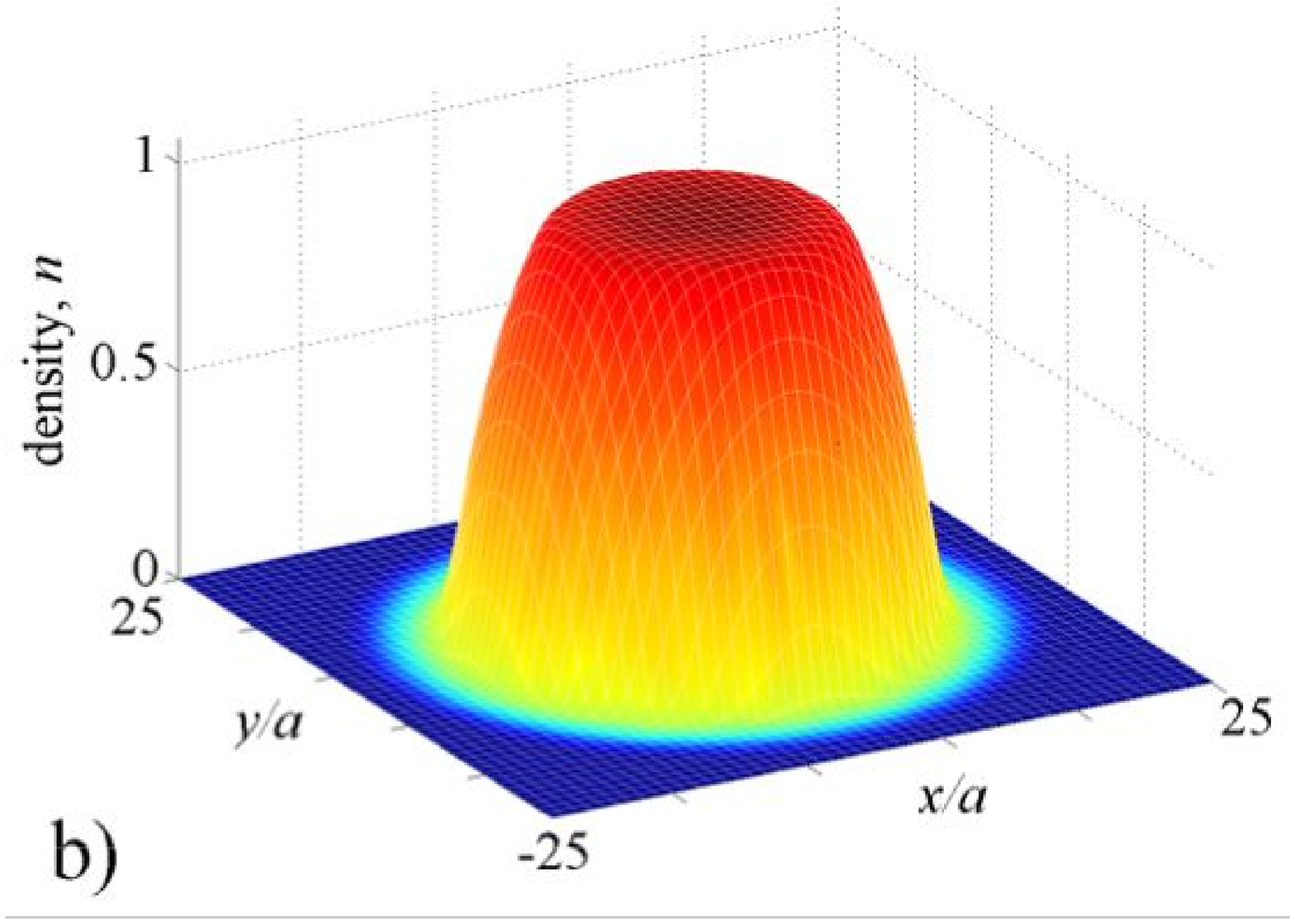}
\caption{ Spatial dependence of the local density for two dimensional
confined atomic boson systems: a) in the superfluid phase; b) at
stronger lattice depths a central Mott insulating region is
formed. The simulations were performed for a parabolic trap with
curvature $V/U=0.002$, at $\mu/U=0.37$, for $U/t=6.7$ (a), and
$U/t=25$ (b).  }
\label{fig:2Dtrap:ldens}
\end{center}
\end{figure}

Next, in Sec.~\ref{sec:limit}, we introduce an effective model
describing the inhomogeneous Bose gas inside the confinement
potential.  This allows us to study the finite size effects arising
from the limited extent of the Bose gas in the trap. Using this
effective model, we discuss the absence of quantum critical behavior
for bosons trapped in confinement potentials.

In a third part, Sec.~\ref{sec:fwhm}, we aim at identifying
experimentally accessible quantities that signal restructurings in the
density distribution of the confined Bose gases. For this purpose, we
monitor the evolution of the momentum distribution function upon
increasing the lattice depth of the optical lattice, similar to the
experimental procedure~\cite{greiner}.  We find that a method put
forward recently~\cite{prokofev} for identifying the phase structure
of the confined Bose gas is not appropriate.  However, by monitoring
the evolution of both the coherence fraction and the full width at
half maximum of the coherent part of the momentum distribution
function we obtain clear signals for restructurings of the density
distributions. Finally, we summarize our observations in
Sec.~\ref{sec:discussion}, along with implications for the experiments
in confined atom systems.

\section{Model and numerical techniques}
\label{sec:model}

Cold confined atomic Bose gases in an optical lattice are direct
experimental realizations of the inhomogeneous Bose Hubbard
Hamiltonian~\cite{zoller}
\begin{eqnarray}
\label{eq:hamiltonian}
H&=&-t\sum_{\langle i, j \rangle} \left(b^{\dagger}_i b_j + h.c.\right)\\
& & + \frac{U}{2}\sum_i n_i(n_i-1) + V \sum_i r^2_i n_i - \mu \sum_i n_i,\nonumber
\end{eqnarray} 
where $b^{\dagger}_i,(b_i)$ denotes the creation (destruction)
operator for bosons at lattice site $i$, located a distance $r_i$ from
the center of the trap, and $n_i=b^{\dagger}_ib_i$ the local density
operator.  The nearest neighbor hopping integral $t$, the onsite
Hubbard repulsion $U$ , and the curvature of the parabolic confinement
potential $V>0$ are tunable parameters in experiments.

After evaporative cooling of the atoms, the experiments are, to a good
approximation, performed at constant particle number.  Similarly, the
chemical potential $\mu$ allows us to control the filling of the trap
in the numerical simulations.  For the discussions below we introduce
a local chemical potential
\begin{equation}
\label{eq:effmu}
\mu^{\rm eff}_i=\mu-Vr^2_i
\end{equation}
which decreases upon moving away from the trap center.

Quantum Monte Carlo (QMC) simulations of the Hamiltonian
Eq.~(\ref{eq:hamiltonian}) were performed using the Stochastic Series
Expansion method~\cite{sse,loopoperator}, with directed loop
updates~\cite{sylju1,sylju2,alet}. This algorithm requires a cutoff
$N_{\rm max}$ on the local site occupation.  Performing simulations at
mean densities $\langle n_i\rangle\le 1$, a cutoff $N_{\rm max}=2$ or
$3$ can be chosen without introducing significant errors.  The
temperature used in the QMC calculations was chosen low enough to be
essentially in the ground state.

All lengths are set in units of the lattice constant $a$, and the
simulation box is taken large enough, to ensure that outside its
boundary the local density $n_i$ vanishes due to depopulation by large
negative values of $\mu^{\rm eff}_i$.  For the values chosen in our
simulations we needed to keep up to 500-site chains in 1D, $50\times
50$ square lattices in 2D and up to $16^3$-site cubes in 3D.

To distinguish the local phases in the inhomogeneous system and to
probe for quantum criticality we define the {\it local
compressibility} at site $i$,
\begin{equation}
\label{eq:kappalocal}
\kappa^{\rm local}_i=\left\langle\frac{\partial n}{\partial 
\mu^{\rm eff}_i}\right\rangle=\beta\left(\langle 
n_i n \rangle - \langle n_i \rangle \langle n \rangle \right),
\end{equation}
by the response of the system's density, n, to a local chemical potential
change at site $i$. Here, $\beta=1/T$ denotes the inverse temperature.
Another way of probing for local properties is to measure
local density fluctuations~\cite{batrouni}.
These are related, by 
\begin{equation}
\label{eq:kappaonsite}
\Delta_i=\frac{1}{\beta}\left\langle\frac{\partial n_i}{\partial \mu^{\rm
eff}_i}\right\rangle=\langle n_i^2\rangle - \langle n_i
\rangle^2,
\end{equation}
to a local density response.
In the homogenous case, the global compressibility
\begin{equation}
\kappa=\left\langle\frac{\partial n}{\partial \mu}\right\rangle=\beta\left(\langle
n^2 \rangle - \langle n \rangle^2 \right)
\end{equation}
is equal to the local comressibility 
$\kappa^{\rm local}_i$ defined in Eq.~(\ref{eq:kappalocal})
at each lattice site $i$, but is in general {\it not} given by $\beta\Delta_i$. 
This is due to correlations between different lattice sites in the superfluid regime. Only 
in the Mott-insulating phases of the homogenous system, where these correlations are absent, 
do all quantities become zero. We will show in Sec.~\ref{sec:lcomp} that the local 
compressibility, $\kappa^{\rm local}$, as defined in Eq.~(\ref{eq:kappalocal}) 
is indeed able to
characterize the local state near a given site even
in the inhomogenous case.

In addition to measuring local quantities, we measure the momentum
distribution function,
\begin{equation}
\label{eq::greensfunction}
n({\bf k})=\frac{1}{N}\sum_{i,j} e^{i({\bf r}_i-{\bf r}_j){\bf k}}
\langle b^\dagger_i b_j \rangle,
\end{equation}
where $N$ denotes the total number of particles within the system.
This way the momentum distribution is normalized to unity, and the
coherence fraction given by the height of the coherence peak, $n({\bf
k}=0)$.

The momentum distribution function $n({\bf k})$ is directly accessible
in experimental studies of confined atomic gases. For negligible
interparticle interactions during ballistic expansion of the atomic
cloud, it essentially maps onto the interference pattern of the
resulting absorption imagines~\cite{roth}.  Due to the tight binding
approximation of the Bose-Hubbard model, the numerical momentum
distribution functions are periodic in the extended zone scheme.  A
finite extent of the Wannier functions on each site in an optical
lattice leads to a form factor in the
momentum distribution~\cite{prokofev}, resulting in unequal heights of
the higher order Bragg peaks, as observed in the experimental
interference patterns~\cite{greiner}.

To further quantify spatial correlations within the trap, we measure
the one particle density matrix, i.e. the correlation function
\begin{equation}
\label{eq:densitymatrix}
g(i,j)=\langle b^\dagger_i b_j \rangle,
\end{equation}
between bosons on lattice sites $i$ and $j$.

\section{Simulations of realistic 2D traps}
\label{sec:sim2D}

\subsection{Phase coexistence in trapped systems}
Trapped inhomogeneous systems can show coexistence of both superfluid
and Mott insulating regions, as has been clearly identified in
experiments~\cite{greiner,esslinger}, in mean field
investigations~\cite{zoller}, numercial renormalization group~\cite{pollet} and 
Gutzwiller ansatz calculations~\cite{schroll}
and in numerical simulations in
1D~\cite{batrouni,kollath,bergkvist} and 3D~\cite{prokofev}.  Simulating large
realistic systems in 2D, we can clearly observe this coexistence. For
large hoppings $t$ the total system is superfluid
(Fig.~\ref{fig:2Dtrap:ldens}a).

As the hopping amplitude $t$ is decreased, a Mott insulating plateau
with integer density (here $\langle n_i\rangle =1$) forms in the
center of the trap, as seen in Fig.~\ref{fig:2Dtrap:ldens}b.  A
superfluid ring-like region with a non-uniform particle density
surrounds the central plateau.  The typical width of this superfluid
ring is $6-8$ times the lattice unit, for the parameters used in our
simulations.  Far away from the center, the density is $\langle
n_i\rangle =0$ which can also be viewed as a Mott plateau.

The emergence of the Mott plateau and the shrinking of the superfluid
region have been interpreted as a quantum phase
transition~\cite{greiner}. We prefer to view it as a crossover where
the volume fraction of the Mott insulating phase grows and that of the
superfluid phase shrinks. After discussing the quantitative properties
of the two phases and of the boundary region we will, in
Sec.~\ref{sec:absence}, show that no quantum critical behavior is
observed in this system, which supports our interpretation of this
phenomenon as a crossover instead of a phase transition.

\subsection{Local compressibility}
\label{sec:lcomp}
\begin{figure}
\begin{center}
\vspace{0.5cm}
\includegraphics[width=8cm]{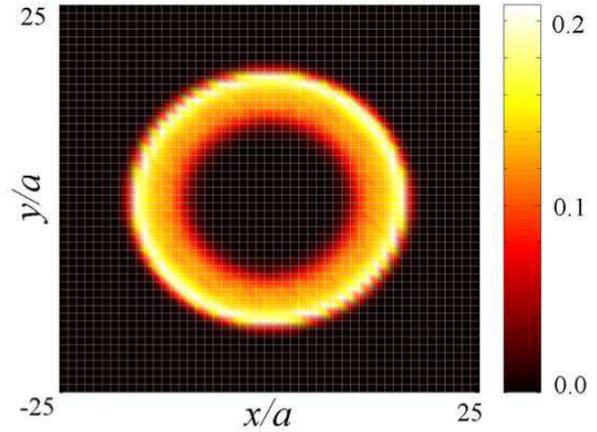}
\caption{Spatial dependence of the local compressibility, $\kappa^{\rm
local}$, of bosons in a two dimensional parabolic trap with curvature
$V/U=0.002$, for $\mu/U=0.37$ and $U/t=25$.  A superfluid ring
surrounding the $n=1$ central Mott plateau is clearly resolved.
}
\label{fig:2Dtrap:kappa2D}
\end{center}
\end{figure}

Fig.~\ref{fig:2Dtrap:ldens} is  useful for a qualitative  view, but we
need a quantitative probe to distinguish the different phases. This
probe is given by the compressibility.  QMC results for the local
compressibility within the trap of Fig.~\ref{fig:2Dtrap:ldens}b are
shown in Fig.~\ref{fig:2Dtrap:kappa2D}. The extent of the superfluid
shell is clearly reflected by this local probe, which also shows that
the compressibility is largest close to the outer regions of the
superfluid shell.

\begin{figure}
\begin{center}
\vspace{0.5cm}
\includegraphics[width=8cm]{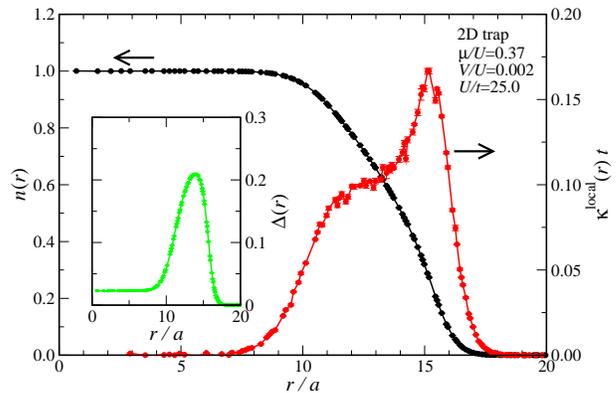}
\caption{ Radial dependence of the density, $n$, and local
compressibility, $\kappa^{\rm local}$, of bosons in a two dimensional
parabolic trap with curvature $V/U=0.002$, for $\mu/U=0.37$ and
$U/t=25$, with a superfluid ring surrounding the $n=1$ central
Mott plateau.  The inset shows the radial dependence of the 
local density fluctuations,
$\Delta$, which remain finite within
the Mott plateau region.
}
\label{fig:2Dtrap:kappaVr}
\end{center}
\end{figure}

For a quantitative analysis we plot in Fig.~\ref{fig:2Dtrap:kappaVr}
the radial dependence of both the local density $n(r)$ and the local
compressibility $\kappa^{\rm local}(r)$ as a function of the distance
$r$ from the center of the trap.  We observe two well defined
Mott regions with local density $n(r)=1$ and $n(r)=0$. In these
Mott plateaux, the local compressibility vanishes, whereas it is
finite in the intervening superfluid ring with non-integral density
$0<n(r)<1$.

More precisely, we observe two well defined peaks in $\kappa^{\rm
local}(r)$, signaling an increase in the particle number fluctuations
at the boundaries to the Mott regions.  While these two peaks would be
of the same height in hard-core boson models (due to particle-hole
symmetry), they are asymmetric in the soft-core case.

The inset of Fig.~\ref{fig:2Dtrap:kappaVr} shows the radial dependence of the local density
fluctuations of Eq.~(\ref{eq:kappaonsite}), $\Delta(r)$, 
first used in
simulations of 1D confined Bose systems~\cite{batrouni}. As opposed to $\kappa^{\rm local}(r)$, 
this
quantity peaks in
the middle of the superfluid ring, and
does not vanish in the central Mott plateau region.
While density fluctuations are therefore most pronounced in 
the middle of 
in the superfluid, they remain finite inside the Mott plateau, due to the 
surrounding superfluid.
These results demonstrate that the local compressibility
$\kappa^{\rm local}(r)$ is a better probe for the existence of superfluid or Mott regions in the
system than the onsite response expressed by $\Delta(r)$.  Moreover, the response of the
{\it total} system to a local excitation should be easier to study experimentally than the local
response to a local excitation (for example by changing the laser intensity at one specific
point).

\subsection{Spatial correlations in the superfluid}
\label{sec:correlations}

\begin{figure}
\begin{center}
\vspace{0.5cm}
\includegraphics[width=8cm]{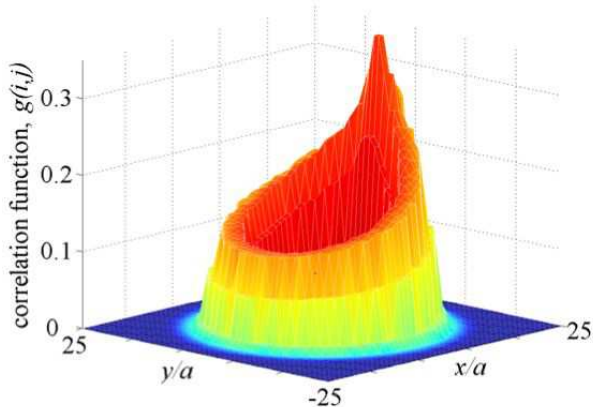}
\caption{ Spatial dependence of the correlation function
$g(i,j)=\langle b^\dagger_ib_j\rangle$ between bosons at site ${\bf
r}_j=(12.3,0)$ inside the superfluid shell, and all other sites $i$
within a two dimensional parabolic trap with curvature $V/U=0.002$, at
$\mu/U=0.37$ and $U/t=25$.
}
\label{fig:2Dtrap:correlations2D}
\end{center}
\end{figure}

In order to gain a better understanding of the extended superfluid
ring, we study the behavior of the one particle density matrix,
Eq.~(\ref{eq:densitymatrix}), in the trapped system.  In
Fig.~\ref{fig:2Dtrap:correlations2D}, the spatial dependence of
$g(i,j)$ is shown for a fixed site $j$, well inside the superfluid
ring, and all other sites $i$ in the parabolic trap.  The rapid decay
of the correlations towards both the central Mott plateau region and
the outside clearly exhibits the ring-like structure of the coherent
superfluid.

\begin{figure}
\begin{center}
\vspace{0.5cm}
\includegraphics[width=8cm]{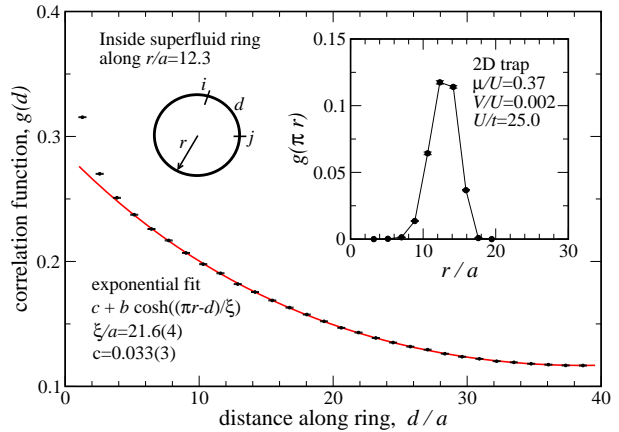}
\caption{ Dependence of the correlation function $g(d)=\langle
b^\dagger_ib_j\rangle$ on the distance $d$ between sites $i$ and $j$
along the superfluid ring for sites within $r/a=12.3\pm1.2$ lattice
units from the center of a two dimensional parabolic trap with
curvature $V/U=0.002$, at $\mu/U=0.37$ and $U/t=25$.  For $d/a>5$ the
data fit well ($\chi^2=1.6$) to an exponential decay, with a finite
value of $c=0.033\pm0.003$, and correlation length
$\xi/a=21.6\pm0.4$. The right inset shows the correlation at the
largest distance, $g(\pi r)$, as a function of the distance $r$ from
the trap center.  The left inset illustrates the distance $d$ between
two sites $i,j$ along a ring of radius $r$ employed here.
}
\label{fig:2Dtrap:correlations1D}
\end{center}
\end{figure}

Inside the superfluid ring the correlation function decays less
rapidly.  In order to make a quantitative analysis of its long
distance behavior, we plot in Fig.~\ref{fig:2Dtrap:correlations1D} the
dependence of the correlation function $g(i,j)$ for site $i,j$ along
the ring on the distance $d$ between the sites, as measured along the
ring. See the left inset of Fig.~\ref{fig:2Dtrap:correlations1D} for
an illustration.  We find all data for distances $d/a>5$ to follow
very closely a finite size exponential decay,
\begin{equation}
g(d)=c+b \cosh\left(\frac{\pi r - d}{\xi}\right),
\end{equation}
with a correlation length $\xi/a=21.6\pm0.4$, and a finite constant
$c=0.033\pm0.003$.  Thus, the superfluid ring does not exhibit
quasi-long ranged correlations, as might have been expected from the
reduced dimensionality of the quasi-1D superfluid ring. Instead, we
find the ring to be wide enough for long range order, {\it i.e.} 2D
behavior, to persist inside the superfluid in spite of the presence of
a central Mott plateau.

A rapid decrease of the correlations, as one moves away from the
middle of the superfluid, can be seen from the right inset of
Fig.~\ref{fig:2Dtrap:correlations1D}, which shows the radial
dependence of the correlation function $g(d)$ at the largest distance
$d=\pi r$.  The correlations are thus most pronounced near $r\approx
12$, which corresponds to the central region of the superfluid ring,
and become suppressed as one moves towards either boundaries.
Finally, inside the Mott plateau region, the correlations are merely
short ranged.  Even though the coherent part of the Bose gas is thus
confined to a 1D ring-like structure, the spatial dependence of the
correlations inside the ring clearly exhibit the underlying 2D
structure of the trapping potential.

Our findings are related to recent results in 1D
systems~\cite{rigol04,kollath}, which show that the presence of a
Mott insulating region does not change the long distance behavior of
the one particle density matrix in trapped boson systems. Similar
analysis might also apply to 3D confinement potentials, which would
allow for extended 2D superfluid shells surrounding a Mott insulating
central region.

\subsection{Local potential approximation}

\begin{figure}
\begin{center}
\vspace{0.5cm}
\includegraphics[width=8cm]{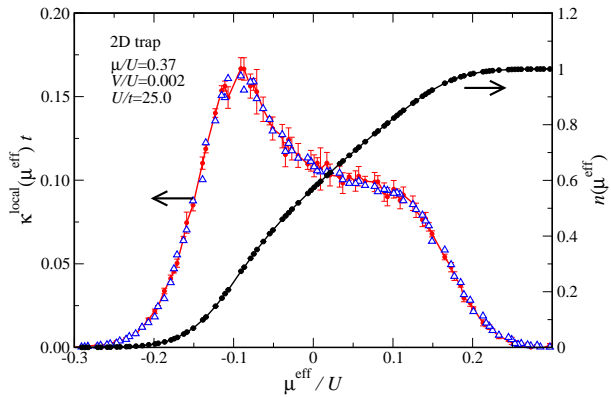}
\caption{ Local density, $n$, and local compressibility, $\kappa^{\rm
local}$, of bosons in a two dimensional parabolic trap with curvature
$V/U=0.002$, at $\mu/U=0.37$ and $U/t=25$, as functions of $\mu^{\rm
eff}$. The results for $\kappa^{\rm local}$ from the density
fluctuations, Eq.(~\ref{eq:kappalocal}) (circles) agree with the
numerical derivative $\partial n/ \partial \mu^{\rm eff}$ (triangles).
}
\label{fig:2Dtrap:kappaVmu}
\end{center}
\end{figure}

The confinement potential $V$ enters the Bose Hubbard Hamiltonian,
Eq.~(\ref{eq:hamiltonian}), by coupling to the boson density.  It
therefore corresponds to a local chemical potential shift, expressed
by the local chemical potential $\mu^{\rm eff}$, defined in
Eq.~(\ref{eq:effmu}).  In Fig.~\ref{fig:2Dtrap:kappaVmu}, we show the
behavior of both the local density and compressibility as functions
of this effective chemical potential, $\mu^{\rm eff}$.  The observed
data collapse in both quantities, as well as the smooth behavior of
these curves for all points on the lattice, indicate that for the
realistic parameters used for the simulation, a local potential
approximation holds, i.e. the local density can be determined from the
value of the local chemical potential. The validity of this
approximation is further confirmed by observing that the local
compressibility coincides perfectly with the numerical derivative of
the curve $n(\mu^{\rm eff})$, i.e.
\begin{equation}
\frac{\partial \langle n\rangle}{\partial \mu^{\rm
eff}_i}=\frac{\partial \langle n_i \rangle}{\partial \mu_0}.
\end{equation}

\begin{figure}
\begin{center}
\vspace{0.5cm}
\includegraphics[width=8cm]{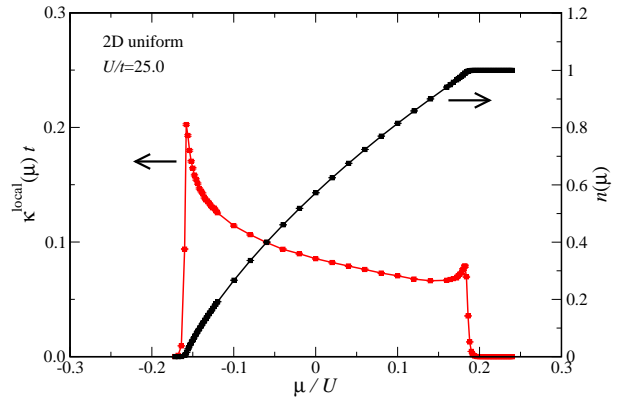}
\caption{ 
Density, $n$, and compressibility $\kappa$ of the
Bose-Hubbard model on the 2D square lattice for $U/t=25$, as a
function of $\mu$. The data are taken from simulations on a $24\times 24$ sites square lattice.
}
\label{fig:2Dtrap:kappaVmu_square}
\end{center}
\end{figure}

It is important to note, that $n(\mu^{\rm eff})$ is not a universal
function, but depends on both the confinement geometry as well as
the trap curvature, $V$.  Universal behavior is violated in
particular near the boundary of the superfluid, where $n(\mu^{\rm
eff})$ does {\it not} behave as $n(\mu)$ for uniform systems,
corresponding to $V=0$.  For example, in the uniform 2D case a cusp
develops when $n$ approaches $1$, as seen in
Fig.~\ref{fig:2Dtrap:kappaVmu_square}, due to quantum
criticality. This cups is absent in the confined system, where $n$
approaches $1$ rather smoothly (c.f. Fig.~\ref{fig:2Dtrap:kappaVmu}).
While a local potential approximation in terms of an effective local
chemical potential holds, the value of $n(\mu^{\rm eff})$ thus {\it
has to be taken from simulations in a trap}, and not from the uniform
system on the underlying lattice.  Approximative schemes such as those
of Ref.~\cite{oliva}, which assume that locally the system can be
mapped onto an unconfined system at the local value of $\mu^{\rm
eff}$, have to be applied with care.  Being reasonable in the bulk of
both the superfluid and the Mott plateau region, such approaches
become unreliable near the interfaces between superfluid and
Mott insulating regions, where differences between the confined and
unconfined case are most pronounced. There QMC simulations are needed
in order to obtain $n(\mu^{\rm eff})$ for a given trap geometry.

\section{Thermodynamic limits and quantum criticality}
\label{sec:limit}

In the previous section we found that in the boundary layer between
Mott plateaux and superfluid regions, the confined system does not
show cusps in the density profile like the uniform system does as a
function of $\mu$ near the quantum critical points.  This is
indication for the absence of quantum criticality in the confined
system.  It could be argued that this is simply due to finite size
effects, but here we show that the situation is indeed more subtle.
For this purpose, we introduce an effective ladder model for bosons
inside inhomogeneous potentials.

\subsection{Effective ladder model for the superfluid ring}

\begin{figure}
\begin{center}
\vspace{0.5cm}
\includegraphics[width=8cm]{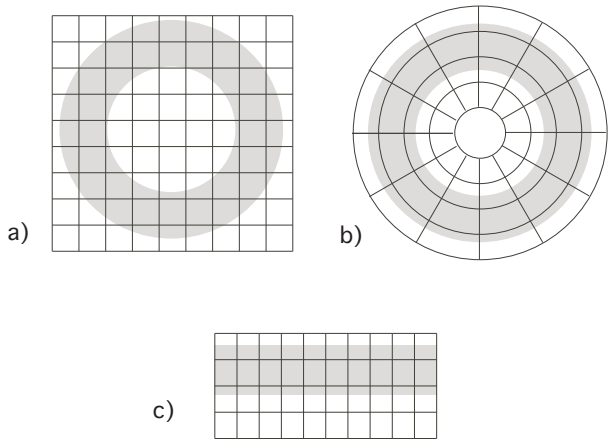}
\caption{ 
Different lattice topologies and expected superfluid regions
(denoted in grey) for a quadratic confining potential for such
topologies: (a) square lattice, (b) polar lattice, (c) ladder lattice.
}
\label{fig:ladder:topology}
\end{center}
\end{figure}

A parabolic potential imposed on a regular hyper-cubic lattice makes
for an irregular form of the superfluid ring surrounding the
Mott insulator (see Fig.~\ref{fig:ladder:topology}a). We thus first
investigate whether this structural quenching is the reason for the
apparent absence of quantum critical features by looking at a
``smoother'' lattice.  Although it cannot be realized in experiments,
we consider a polar lattice, such as depicted in
Fig.~\ref{fig:ladder:topology}, which is topologically equivalent to
the ladder system in Fig.~\ref{fig:ladder:topology}.  The rotational
invariance of the polar lattice (translation invariance of the ladders
along the leg direction) simplifies the detection of critical behavior
by reducing finite size effects.

While the polar and ladder lattices are not topologically equivalent
to the square lattice, universality of quantum critical behavior
ensures that we study the same effects. Comparing the lattices will be
a further check for the validity of a local potential approximation
for the confined system.

The Hamiltonian of the ladder system of size  $L_x\times L_y$ is
\begin{eqnarray}
\label{eq:ladder}
H&=&-t\sum_{i=1}^{L_x}\sum_{j=1}^{L_y} \left(b^{\dagger}_{i,j}
b_{i+1,j} + b^{\dagger}_{i,j} b_{i,j+1} + h.c.\right)\\ & & +
\frac{U}{2}\sum_{i=1}^{L_x}\sum_{j=1}^{L_y}n_{i,j}(n_{i,j}-1) -
\sum_{j=1}^{L_y} \mu(j) \sum_{i=1}^{Lx} n_{i,j}, \nonumber
\end{eqnarray} 
where the confinement potential is now given by a chemical potential
$\mu(j)$ which is constant for each leg $j$ of the ladder. The
boundary conditions are chosen to be periodic along the legs of the
ladder, and open in the transverse direction.

With these choices, the local density $\langle n_{i,j} \rangle$ is
independent of $i$ due to translation invariance along a leg of the
ladder, and we denote by $n(j)$ the density of particles on the $j$-th
leg.  Similarly, the local compressibility $\kappa^{\rm local}(j)$ is
the same for all sites on the $j$-th leg. The results shown below have
been obtained for a geometry with $L_x=64$, and $L_y=10$.

To simplify simulations, we linearize the quadratic potential,
$\mu+Vj^2$ and set
\begin{equation}
\mu(j)=\mu_0+j\Delta \mu ,
\end{equation}
where $\Delta \mu$ is the difference in chemical potential between two
neighboring legs. In our simulations we fix $\Delta \mu/U=0.053$ and
make sure that the first leg is always in the $\langle n\rangle=0$
Mott insulating phase and the last leg in the $\langle n\rangle =1$
insulating phase. Sweeping the value of the global $\mu_0$ allows us
to obtain results for all values of the local chemical potentials, and
to drive one of the legs across the transition from the superfluid to
the Mott insulating phase.  We note, that a similar construction is
obtained for 3D traps, by modeling the superfluid shell as a set of
coupled planes with a chemical potential gradient applied
perpendicular to the planes.

\subsection{Results of the ladder model and comparison to the realistic trap}

\begin{figure}
\begin{center}
\vspace{0.5cm} \includegraphics[width=8cm]{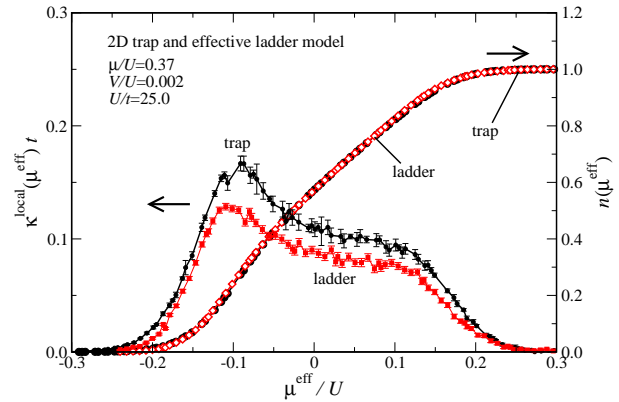}
\caption[]{ Local density, $n$, and local compressibility,
$\kappa^{\rm local}$, for a two dimensional parabolic trap on a square
lattice and for a ladder, as functions of $\mu^{\rm eff}$.  The
parameters for the trapping potential are $V/U=0.002$, at
$\mu/U=0.37$, and the parameters of the ladder model are chosen as to
cover the whole superfluid region (see text). The density profile
within the trap and the ladder coincide almost perfectly.  Small
differences are exhibited by the two different compressibility curves,
which nevertheless share the same overall shape.  }
\label{fig:ladder:lcomp}
\end{center}
\end{figure}

In Fig.~\ref{fig:ladder:lcomp} we show the combined results of all
simulations with different values of $\mu_0$. We plot both the local
density $\langle n(j) \rangle$ and local compressibility $\kappa^{\rm
local}(j)$ for all legs and all values of $\mu_0$ as functions of the
local chemical potential. Note that all data from simulations using
different values of $\mu_0$ collapse onto a single curve.  Due to the
equivalence of all sites on the ladder with the same value of the
chemical potential, this data collapse is expected in the ladder
model, where changing $\mu_0$ by an amount of $\Delta\mu$ corresponds
to a shift in the index of each chain in the ladder.

Fig.~\ref{fig:ladder:lcomp} also shows the results obtained for the
realistic trap, as described in the previous section. Both results
{\it coincide almost perfectly}, which is a strong justification for
the validity of the effective model and the applicability of a local
potential approximation.

The small differences can be attributed to the different shape of the
trapping potential which is parabolic for the realistic trap, but
linearized in the ladder model.  While differences in the densities
are small, the compressibility, which is the derivative of the
density, is more sensitive, marking the lack of universality.

In particular, we find that the ladder model does not show cusp-like
features in the boundary regions between the superfluid and the
Mott plateaux, similar to the realistic 2D trap.  We thus conclude,
that structural quenching in the rotational symmetric trap on the
square lattice is not the reason for the apparent absence of quantum
criticality since similar behavior is also found on the polar lattice,
i.e. in the ladder model.  We expect similar features also to be
obtained for 3D traps using the effective multilayer model.

\subsection{Quantum criticality}
\label{sec:absence}

Having shown that the ladder model captures important features of the
realistic trap even quantitatively, we now discuss possible quantum
criticality in confined systems.

Performing the thermodynamic limit in the uniform case does not change
the model parameters apart from increasing the number of lattice
sites. On the other hand, standard thermodynamic limit definitions for
the confined case~\cite{damle, muramatsu} imply a decrease in the
trapping potential's curvature $V$, such that
\begin{equation}
\label{eq:tdl}
N\rightarrow\infty, \quad {\rm with}\quad  N \sqrt{V/t}={\rm const.}
\end{equation}
Such a procedure locally drives the system towards the uniform regime,
since for lattice sites with a given value of $\mu^{\rm eff}$, the
gradient in the confinement potential eventually becomes irrelevant on
the scale set by the correlation length in their neighborhood.
Therefore, the state of the uniform system at the value of $\mu^{\rm
eff}$ is established there~\cite{damle}.

\begin{figure}
\begin{center}
\vspace{0.5cm} \includegraphics[width=8cm]{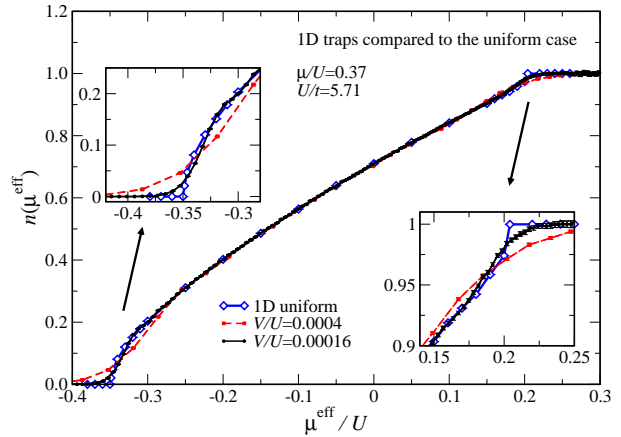}
\caption[]{ Local density, $n$, for bosons confined in one dimensional
parabolic confinement potentials of different curvatures $V$, as a function of
$\mu^{\rm eff}$. For comparison, the behavior of the density as a
function of $\mu$ in the uniform 1D system is also shown. The insets focus on the regions close 
to $n=0$ and $n=1$, where differences between the trapped case and the uniform case are most 
pronounced.} \label{fig:1Dtrap:TDL}
\end{center}
\end{figure}

An example of this approach to the uniform system upon decreasing the
trapping potential $V$ is given in Fig.~\ref{fig:1Dtrap:TDL}. It shows
the density $n$ as a function of $\mu^{\rm eff}$ for two different 1D
traps. In the 1D case, we can study larger systems than in 2D,
allowing us to decrease $V$ by a factor of $25$, which requires
increasing the linear system size by a factor of $5$.  While in both
traps in Fig.~\ref{fig:1Dtrap:TDL} the value of the local density
follows the curve $n(\mu)$ of the uniform 1D system, deviations are
visible in the insets, which focus on the regions close to the
Mott plateaux.  In the case of the shallow trap, these deviations are
clearly reduced, with more pronounced singularities developing at the
critical point.

The thermodynamic limiting procedure in Eq.~(\ref{eq:tdl}) corresponds
to the limit $\Delta\mu\rightarrow 0$ in the gradient of the effective
ladder model.  In this limit, we recover the Bose Hubbard model on the
2D square lattice, showing 2D quantum critical behavior.  A finite
gradient $\Delta\mu$ in the ladder model restricts the correlation
length perpendicular to the chains, and does not allow for 2D quantum
critical behavior to be observed.

\begin{figure}
\begin{center}
\vspace{0.5cm}
\includegraphics[width=8cm]{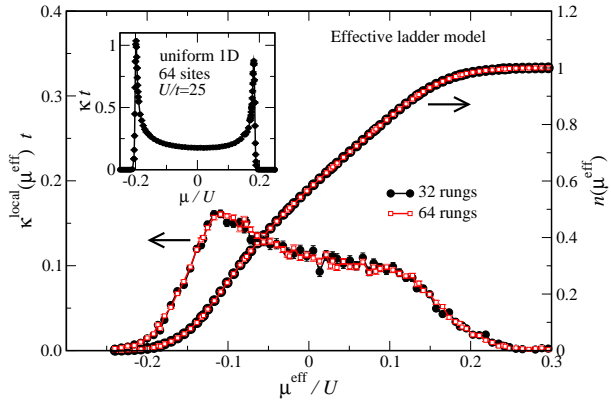}
\caption[]{ Local density, $n$, and local compressibility,
$\kappa^{\rm local}$, for the Bose Hubbard model on ladders with
different lengths as a function of $\mu^{\rm eff}$.  The other
parameters of the ladder model are chosen as in
Fig.~\ref{fig:ladder:lcomp}.  The inset shows the compressibility for
the uniform 1D case as a function of $\mu$ for a chain with 64 sites,
for the same value of $U/t$ as used in the ladder model.  }
\label{fig:ladder:lcompsize}
\end{center}
\end{figure}

However, in the opposite limit, $\Delta\mu\rightarrow \infty$, we
recover a 1D chain on the lowest leg, which will show 1D critical
behavior.  In the inset of Fig.~\ref{fig:ladder:lcompsize} the
compressibility of the Bose Hubbard model is shown on a chain of 64
sites.  Already on such a short chain sharp peaks near the transition
are visible, which diverge with increasing chain
length~\cite{batrouni_kappa}.

One might expect this 1D quantum criticality to persist in the ladder
model with a finite gradient $\Delta\mu$, observed upon changing the
chemical potential $\mu_0$ to drive one of the legs across the phase
transition.
This is however not the case: While the broad peaks in
Fig.~\ref{fig:ladder:lcomp} are weak remnants of 1D quantum
criticality, they do {\it not} diverge as the size of the ladder model is
increased. This is seen from Fig.~\ref{fig:ladder:lcompsize}, which
shows the density and local compressibility 
as functions of $\mu^{\rm eff}$
for ladders of different
lengths.

Therefore, the effective ladder model as well as the realistic trapped
system {\it do not show quantum criticality}.  The quantum critical
behavior that might have been expected to occur in the boundary layer
between the Mott insulator and the superfluid is destroyed by the
finite gradient in the effective chemical potential, and the coupling
of this layer to the rest of the system.
The absence of quantum criticality in the realistic 2D parabolic 
trap is not
only due to its finite size, but is in fact imposed by the
inhomogeneity of the confinement potential, as we showed using
the effective ladder description.  The Mott transition observed in
experiments on ultra-cold atoms~\cite{greiner} should not be viewed as
a quantum phase transition, but instead as a crossover with changing
volume fractions of the Mott plateaux and superfluid regions.

In contrast, for flat confinement potentials, corresponding to $V=0$,
quantum critical behavior can be observed already on 
moderate system sizes. For example, the data shown in
Fig.~\ref{fig:2Dtrap:kappaVmu_square} for a $24\times 24$ sites
2D system, and in the inset of Fig.~\ref{fig:ladder:lcompsize},
for a 64 sites chain, both clearly reflect the quantum
criticality of the uniform 2D and 1D cases.

\section{Identifying Mott plateau formation in traps}
\label{sec:fwhm}

While measuring the spatial density distribution and the local
compressibility in QMC simulations allows for the identification of
the different regions inside the trap, such a local probe is not (yet)
available experimentally.

In order to identify changes in the density distribution inside the
trap upon varying control parameters such as $V$, $U$, or $t$,
different strategies have been proposed. In particular, in
Ref.~\cite{prokofev} it was claimed that the presence of a
Mott plateau in the trap center is signaled by fine structures in the
momentum distribution function, and that these are absent in traps
without Mott plateaux.

\begin{figure}
\begin{center}
\vspace{0.5cm}
\includegraphics[width=8cm]{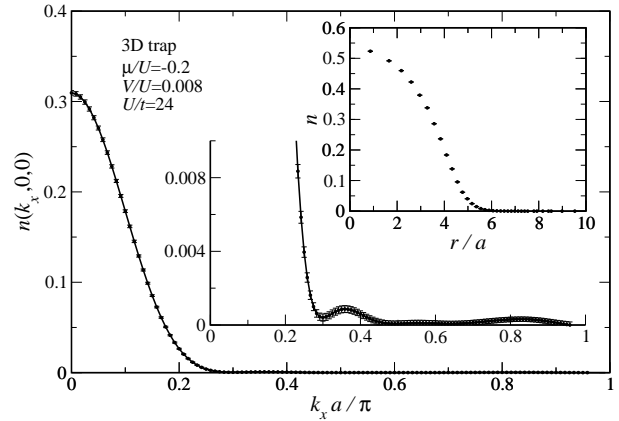}
\caption{ Momentum distribution function of bosons in a
three dimensional parabolic trap with curvature $V/U=0.008$, along the
line $(0,0,0)$-$(\pi/a,0,0)$ in momentum space, for $\mu/U=-0.2$, and
$U/t=24$, values taken from Ref.~\cite{prokofev}.  The lower inset
exhibits the presence of satellite peaks in this system, without a
Mott plateau, as seen from the radial density distribution shown in
the top inset.
}
\label{fig:3Dprokofev:a}
\end{center}
\end{figure}

\begin{figure}
\begin{center}
\vspace{0.5cm}
\includegraphics[width=8cm]{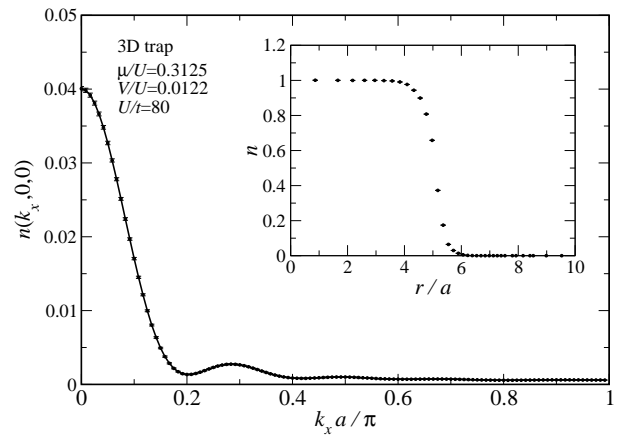}
\caption{ Momentum distribution function of bosons in a
three dimensional parabolic trap with curvature $V/U=0.0122$, along
the line $(0,0,0)$-$(\pi/a,0,0)$ in momentum space, for
$\mu/U=-0.3125$, and $U/t=80$, values taken from
Ref.~\cite{prokofev}. The corresponding radial density distribution is
shown in the inset.
}
\label{fig:3Dprokofev:c}
\end{center}
\end{figure}

Two of the configurations of Ref.~\cite{prokofev} are shown in
Figs.~\ref{fig:3Dprokofev:a} and \ref{fig:3Dprokofev:c}. Here, we
simulated 3D systems in a trap and used the parameters given in
Ref.~\cite{prokofev}, and plot the resulting momentum distribution
along the line from ${\bf k}=(0,0,0)$ to $(\pi/a,0,0)$, along the
$k_x$-axis of momentum space.

While in Fig.~\ref{fig:3Dprokofev:c}, where an extended Mott plateau
is present, additional fine structures in $n({\bf k})$ are visible,
such structures seem not to exist at first sight for the data shown in
Fig.~\ref{fig:3Dprokofev:a}.  However, similar fine structures are
also present there, even on a similar scale, as clearly seen from the
bottom inset of Fig.~\ref{fig:3Dprokofev:a}, which focuses on the tail
of the momentum distribution function.  These fine structures become
invisible on the scale used for the main part of
Fig.~\ref{fig:3Dprokofev:a}, due to the presence of a dominant
coherence peak.  We found in the simulations discussed below, that a
simple correspondence as proposed in Ref.~\cite{prokofev} does not
hold: The presence of fine structures is due to the finite extent of
the superfluid within the trap, and emerges also without a
Mott plateau being present in the trap center.  Denoting the radial
extent of the superfluid by $R$, these peaks appear for $k_x$ near
integer multiplies of $2\pi/R$, if not masked by the incoherent
background, or rendered almost invisible on the scale of the coherence
peak.

For a more systematic analysis of the experimentally accessible
momentum distribution function we monitor its evolution upon varying
experimentally accessible control parameters of the system.  We have
performed sets of simulations, which mimic the experimental procedure
of increasing the lattice depth of the optical lattice~\cite{greiner},
by reducing the hopping parameter $t$, while keeping all other
parameters constant.  We performed such scans for traps of different
dimensionalities, and begin our discussion of the results for the 2D
case.

\subsection{Homogeneous and closed box 2D systems}

In order to better interpret data obtained for inhomogeneous traps, we
first consider the homogeneous case.  To this end, we set the trap
curvature $V=0$, using periodic boundary conditions (PBC), and also
consider a finite square lattice with open boundary conditions (OBC),
representing an infinitely sharp trapping potential, i.e. a closed box
system.  We then compare both the closed box and the parabolic trap to
the uniform case.

\begin{figure}
\begin{center}
\vspace{0.5cm}
\includegraphics[width=8cm]{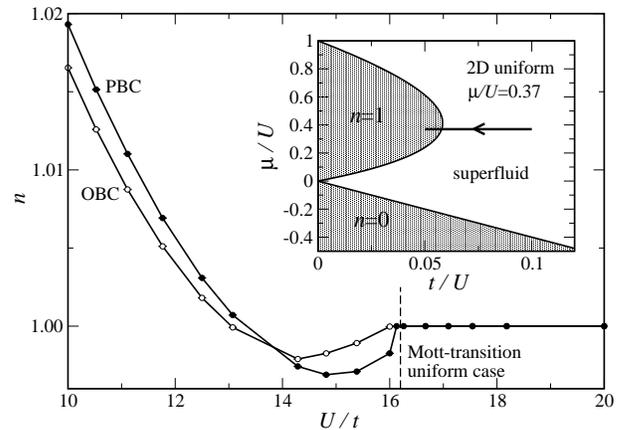}
\caption{ Density $n$ as a function of $U/t$ for bosons along a
constant $\mu/U=0.37$ scan, for both periodic (PBC) and open (OBC)
boundary conditions on a 24$\times$24 sites square lattice, OBC
corresponding to an infinitely sharp trapping potential.  The
Mott transition in the uniform case is indicated by the dashed line.
The inset locates this constant $\mu$ scan within the phase diagram of
the uniform system.
}
\label{fig:2Dhomo:density}
\end{center}
\end{figure}

The phase diagram of the Bose Hubbard model on the square lattice in
the parameter regime of interest is shown in the inset of
Fig.~\ref{fig:2Dhomo:density}.  While the overall phase structure is
well described by mean field theory~\cite{fisher,monien}, the maximum
extent of the first Mott lobe of $U/t=23.3$ is thereby overestimated
by about 32\% of the value obtained by a strong coupling expansion
($U/t=17.54$)~\cite{monien,freericks}.  Thus, in order to compare
later the onset of Mott plateau formation in confined systems with the
homogeneous case, we first need to determine the precise phase boundary
within the region of interest.

In the following we consider a path through the phase diagram along a
line of constant $\mu/U=0.37$, indicated by the directed line in the
inset of Fig.~\ref{fig:2Dhomo:density}.  Moving along this line
through the quantum phase transition, the system undergoes the generic
transition~\cite{fisher} which is mean field in nature. The transition
does not cross over to the special case of the commensurate
transition, belonging to the $3D$ $XY$ universality
class~\cite{fisher}.

\begin{figure}
\begin{center}
\vspace{0.5cm}
\includegraphics[width=8cm]{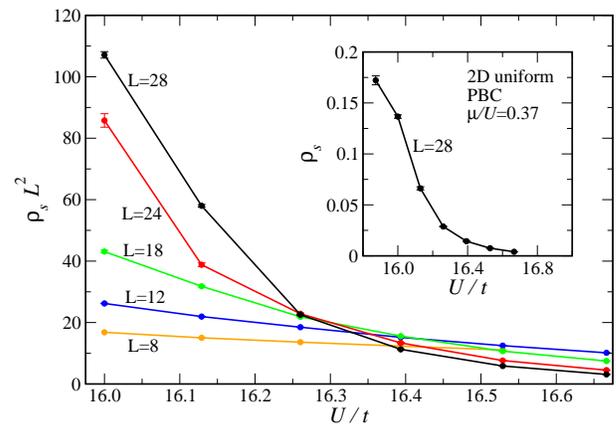}
\caption{ Scaling plot of the stiffness, $\rho_s$, for the Bose
Hubbard model on the square lattice for periodic boundary conditions,
while tuning along a constant $\mu/U=0.37$ scan. The inset shows the
stiffness, $\rho_s$, for a $28 \times 28$ sites system as a function
of $U/t$.
}
\label{fig:2Dhomo:stiffness}
\end{center}
\end{figure}

In order to identify the Mott transition point along this scan we
measured the stiffness, $\rho_s$, which quantifies the response of the
system to a twist in the boundary conditions along the real space
directions. This quantity can be calculated in QMC from the boson
winding number fluctuations~\cite{winding}. At the quantum critical
point finite size scaling theory~\cite{fisher} predicts it to scale in
two dimensions as $L^{-z}$, where $z=2$ is the dynamical critical
exponent for the generic transition, and $L$ the linear system size of
a $N=L^2$ sites system.

The inset of Fig.~\ref{fig:2Dhomo:stiffness} shows $\rho_s$ as a
function of the inverse hopping, $U/t$, for a system of linear size
$L=28$, indicating its increase upon entering the superfluid phase.
The main part of Fig.~\ref{fig:2Dhomo:stiffness} is a scaling plot of
$\rho_s L^2$ vs. $U/t$, from which the critical point
$U/t=16.25\pm0.1$ is obtained with sufficient precision for the
purpose of this study.  Note that corrections to scaling are visible
already at $L=12$, in agreement with a recent study, indicating that
large systems sizes are needed for high precision determinations of
the critical point~\cite{aletgeneric}.  Mean field
theory~\cite{monien} predicts a value of~$U/t=23.5$ which is too
large by more than 40\%.  Thus in 2D, corrections to mean field have to
be accounted for already the uniform case.
 
Having established the position of the Mott transition in the uniform
case, we show in Fig.~\ref{fig:2Dhomo:density} the evolution of the
density, $n$, in homogeneous systems with PBC and OBC along the
constant $\mu$ scan shown in the inset of
Fig.~\ref{fig:2Dhomo:density}. Here, we show data on systems with $24\times 24$ lattice sites.
In both cases the onset of the
Mott insulating regime is located close to the critical point obtained
from finite size scaling. Thus already for this moderate system size,
the Mott transition is located rather close to the value in the thermodynamic limit.

\begin{figure}
\begin{center}
\vspace{0.5cm}
\includegraphics[width=8cm]{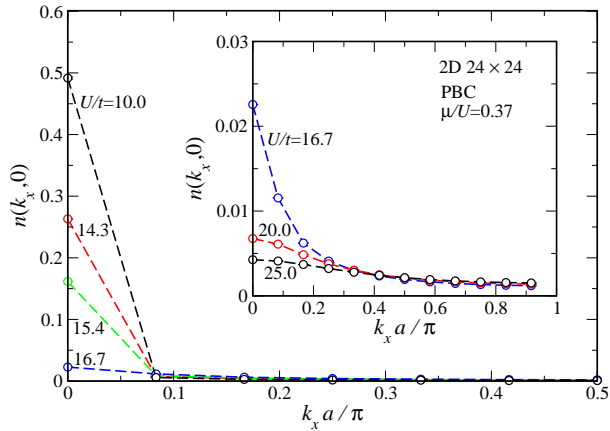}
\caption{ Momentum distribution function of bosons on a 24$\times$24
sites square lattice with periodic boundary conditions (PBC) for the
commensurate momenta along the line $(0,0)$-$(\pi/a,0)$ in momentum
space, for different values of $U/t$, while tuning $t$ along the
constant $\mu/U=0.37$ scan of Fig.~\ref{fig:2Dhomo:density}.  The loss
of coherence due to the Mott transition at $U/t=16.7$ is reflected by
the reduced coherence peak height, $n(0,0)$.
}
\label{fig:2Dhomo:green1}
\end{center}
\end{figure}

\begin{figure}
\begin{center}
\vspace{0.5cm}
\includegraphics[width=8cm]{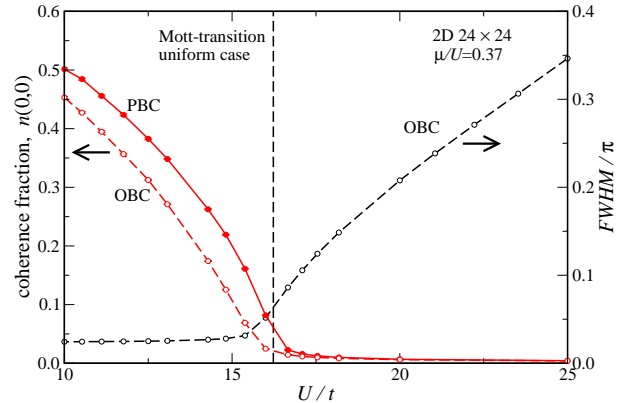}
\caption{ Evolution of the coherence fraction, i.e. the height of the
coherence peak, $n(0,0)$, as a function of $U/t$ while tuning $t$
along the constant $\mu/U=0.37$ scan of Fig.~\ref{fig:2Dhomo:density},
for bosons on a $24\times 24$ sites square lattice for both periodic (PBC) and open (OBC) 
boundary conditions. For OBC
the full width at half maximum ($FWHM$) of the coherence peak is also
shown, clearly signaling the onset of the Mott phase. The
Mott transition point is indicated by the vertical dashed line.
}
\label{fig:2Dhomo:fwhm}
\end{center}
\end{figure}

Phase coherence is signaled by a pronounced coherence peak in the
momentum distribution function, $n({\bf k}=0)$.  This can be seen from
Fig.~\ref{fig:2Dhomo:green1}, which shows $n({\bf k})$ for the
commensurate momenta along the $k_x$-axis of momentum space for
different values of $U/t$ for a $24 \times 24$ sites system with
PBC. The evolution of $n(0)$ while tuning through the transition is
shown in Fig.~\ref{fig:2Dhomo:fwhm}. It clearly marks the loss of
coherence in the Mott insulator.

\begin{figure}
\begin{center}
\vspace{0.5cm}
\includegraphics[width=8cm]{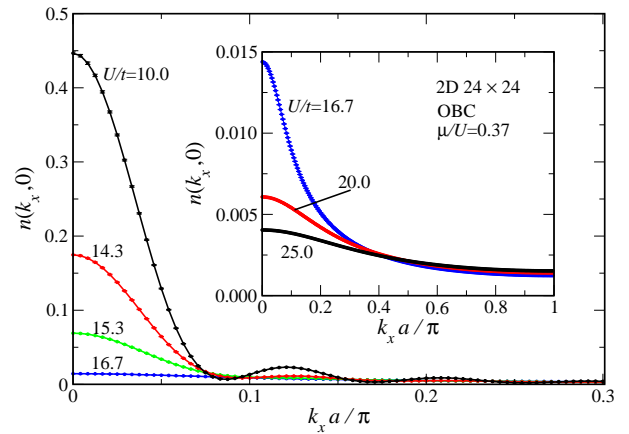}
\caption{ Momentum distribution function of bosons on a
24$\times$24-site square lattice with open boundary conditions,
corresponding to an infinitely sharp trapping potential, along the
line $(0,0)$-$(\pi/a,0)$ in momentum space, for different values of
$U/t$, while tuning $t$ along the constant $\mu/U=0.37$ scan of
Fig.~\ref{fig:2Dhomo:density}. Within the superfluid regime satellite
peaks appear, which diminish upon emergence of the Mott phase.
}
\label{fig:2Dhomo:green0}
\end{center}
\end{figure}

While for PBC a discrete set of commensurate momenta exists, for OBC
states of arbitrary momenta can be occupied.  In the QMC simulations,
we measured the momentum distribution function on a mesh covering
$10\times L$ momenta in the first Brillouin zone along the
$k_x$-axis. The resulting momentum distribution functions for
different values of $U/t$ are shown in
Fig.~\ref{fig:2Dhomo:green0}. Similar to the case of PBC, the loss of
coherence upon entering the Mott phase is signaled by a reduction of
the coherence fraction, $n(0)$, as seen from
Fig.~\ref{fig:2Dhomo:fwhm}.  Also note the pronounced fine structures
in $n({\bf k})$ for $U/t=10.0$, deep in the superfluid regime, and the
complete absence of such structures for larger values of $U/t$.

Using a Ornstein-Zernike form for the coherent part for the momentum
distribution function,
\begin{equation}
n({\bf k})=\frac{n(0)}{1+{\bf k}^2\xi^2},
\end{equation}
where $\xi$ denotes the coherence length, we can obtain $\xi$ from the
full width at half maximum ($FWHM$) of the coherence peak,
\begin{equation}
FWHM=\frac{2}{\xi}.
\end{equation}

In the thermodynamic limit, the coherence length diverges in the
superfluid regime.  On a finite system it is bounded from above by the
linear system size $L$, and decreases to zero deep inside the
Mott insulating regime. We thus expect an increase of the $FWHM$, from
its minimum value of $2/L$ in the superfluid regime, upon driving the
system through the Mott transition.  This behavior can indeed be seen
in Fig.~\ref{fig:2Dhomo:fwhm}, which shows the $FWHM$ as a function of
$U/t$ for the $24\times 24$ sites closed box. The $FWHM$ is at its
lowest value of about $2/24$ left of the Mott transition, and
increases due to loss of coherence beyond this point.

\subsection{Parabolic traps in 2D}

\begin{figure}
\begin{center}
\vspace{0.5cm}
\includegraphics[width=8cm]{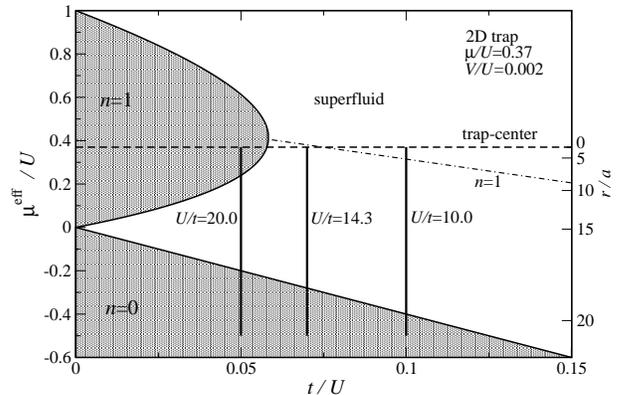}
\caption{ Variation of $\mu^{\rm eff}$ in a two dimensional parabolic
trap with curvature $V/U=0.002$ for three different values of $U/t$
while tuning $t$ along a constant $\mu/U=0.37$ scan, shown within the
phase diagram of the Bose Hubbard model on the square lattice. Along
the dashed dotted line the density has a constant value of $n=1$.
}
\label{fig:2Dtrap:phasediag}
\end{center}
\end{figure}

After our analysis of the homogeneous system, we are now in position to
discuss the evolution of a confined Bose gas in a 2D optical lattice
while increasing the lattice depth, i.e. decreasing the hopping,
$t$. In the following, we will take the system along a path of
constant chemical potential, $\mu/U=0.37$, in a parabolic trap of
curvature $V/U=0.002$.  An illustration of the path taken in our
simulations is shown in Fig.~\ref{fig:2Dtrap:phasediag}.

As discussed in Sec.~\ref{sec:sim2D}, we can use a local potential
approximation for a qualitative description of the inhomogeneous
density profile of the trap, by employing the local value of $\mu^{\rm
eff}$.  For a given value of $U/t$, we can therefore represent the
confined system by a vertical line in the phase diagram of the Bose
Hubbard model on the square lattice.  This representative vertical
line is then shifted towards smaller values of $t$, during our
constant $\mu$ scan.  While for large values of $t$ the system will be
superfluid, from Fig.~\ref{fig:2Dtrap:phasediag} we expect the
appearance of a central Mott plateau for smaller hopping amplitudes.

\begin{figure}
\begin{center}
\vspace{0.5cm}
\includegraphics[width=8cm]{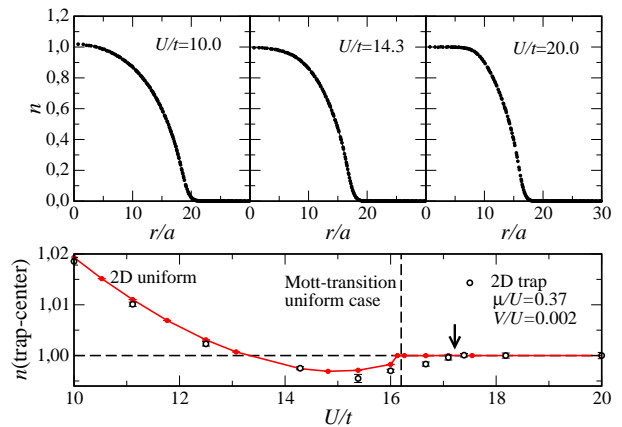}
\caption{ Top: Radial density distribution of bosons within a
two dimensional parabolic trap with curvature $V/U=0.002$ for
different values of $U/t$ for $\mu/U=0.37$.  Bottom: Evolution of the
density in the trap center while tuning $t$ along a
constant $\mu/U=0.37$ scan.  For comparison the density of the uniform
system along the same constant $\mu$ scan is shown. The arrow
indicates the threshold for Mott plateau formation within the trap,
which deviates by about 6\% from the position of the Mott transition
in the uniform case (vertical dashed line).
}
\label{fig:2Dtrap:density}
\end{center}
\end{figure}

In order to determine accurately the position of the Mott plateau
formation, we monitor the evolution of the density in the trap center,
which is shown in Fig.~\ref{fig:2Dtrap:density} as a function of the
inverse hopping, $U/t$.

In accordance with Fig.~\ref{fig:2Dtrap:phasediag}, the trap center
has a density larger than 1 for small values of $U/t$. Upon increasing
$U/t$, the central density decreases, closely following the path of
the homogeneous system. It crosses the value 1 for $U/t\approx13.13$,
as expected from Fig.~\ref{fig:2Dtrap:phasediag}, then undergoes a
minimum, and reaches 1 for $U/t\approx16.7$, where a Mott plateau
starts to form.  Apart from the critical region of the homogeneous 2D
system, the density in the center of the trap closely follows the
behavior in the uniform case along the same constant $\mu$ scan.  The
rather smooth approach towards a central density of $1$ is in
agreement with the observations in Sec.~\ref{sec:absence} of the
absence of critical fluctuations inside the trap.  In addition to this
qualitative difference to the homogeneous case, the local potential
approximation underestimates the lower value of $U/t$ for Mott plateau
formation by about 6\%, as seen from Fig.~\ref{fig:2Dtrap:phasediag}.

Having determined the value of $U/t$ for the onset of the central
Mott plateau from measurements of the density distribution, we turn to
a discussion of the momentum distribution function of bosons inside
the parabolic trap and its evolution upon increasing the inverse
hopping, $U/t$.

\begin{figure}
\begin{center}
\vspace{0.5cm}
\includegraphics[width=8cm]{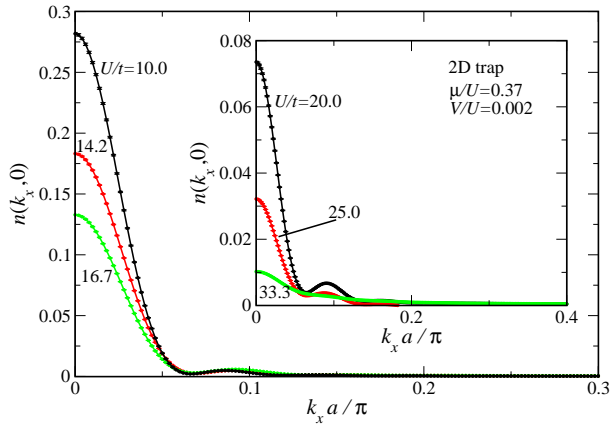}
\caption{ Momentum distribution function of bosons in a
two dimensional parabolic trap with curvature $V/U=0.002$ along the
line $(0,0)$-$(\pi/a,0)$ in momentum space, for different values of
$U/t$ while tuning $t$ along the constant $\mu/U=0.37$ scan of
Fig.~\ref{fig:2Dtrap:density}. Satellite peaks near $(k_x,k_y)=(0.1
\pi/a,0)$ appear unrelated to the phase structure within the trap.
}
\label{fig:2Dtrap:green}
\end{center}
\end{figure}

In Fig.~\ref{fig:2Dtrap:green} the momentum distribution function
$n({\bf k})$ is shown for different values of $U/t$ along the
constant $\mu/U=0.37$ scan of Fig.~\ref{fig:2Dtrap:density}. From
these data it is obvious that the presence of the fine structure peak
near $k_x a /\pi\approx0.1$, reflecting the typical radial extent,
$R\approx 20 a$, of the confined Bose gas, is not related to the
presence of a central Mott plateau, in agreement with earlier
observations in Sec.~\ref{sec:fwhm}.

\begin{figure}
\begin{center}
\vspace{0.5cm}
\includegraphics[width=8cm]{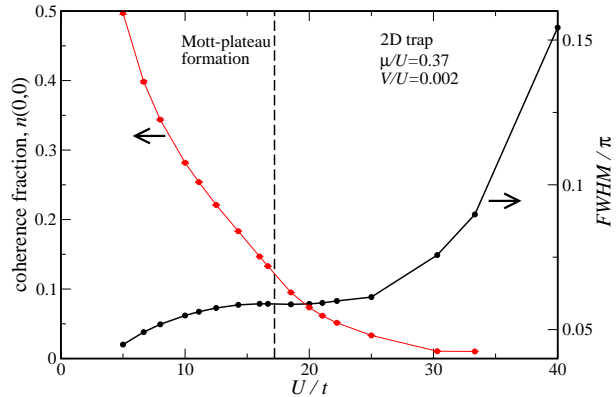}
\caption{ Evolution of the coherence fraction, i.e. the height,
$n(0,0)$, and of the full width at half maximum ($FWHM$) of the
coherence peak as a function of $U/t$ while tuning $t$ along the
constant $\mu/U=0.37$ scan of Fig.~\ref{fig:2Dtrap:density} for bosons
in a two dimensional parabolic trap with curvature $V/U=0.002$. The
threshold for Mott plateau formation is indicated by the dashed line.
}
\label{fig:2Dtrap:fwhm}
\end{center}
\end{figure}

Analyzing the data further, we show in Fig.~\ref{fig:2Dtrap:fwhm} the
coherence fraction, $n({\bf k}=0)$, and the $FWHM$ of the coherence
peak as a function of $U/t$.  In marked contrast to the uniform case
(Fig.~\ref{fig:2Dhomo:fwhm}), but in agreement with experimental
findings~\cite{greiner}, the coherence fraction does not display
distinct features upon emergence of a central Mott plateau, but
instead decreases rather smoothly over a broader range than for the
uniform system.  This behavior is expected as it reflects the
coexistence of both a Mott plateau region and a surrounding
superfluid.

Similar broadenings are observed in the evolution of the $FWHM$, which
becomes rather flat in the region where Mott plateau formation sets
in.  Since changes in the $FWHM$ are thus small near the threshold,
care has to be taken when extrapolating data taken for large values of
$U/t$ down to the flat region.  For example, a linear extrapolation
using the last two data points in Fig.~\ref{fig:2Dtrap:fwhm}, would
overestimate the threshold for Mott plateau formation by more than
60\% of its actual value.

However, we find that the $FWHM$ starts to increase well below the
threshold for Mott plateau formation. Furthermore, beyond this point a
change in curvature of the graph is observed, with an inflection point
located at the threshold point.  In fact, we found the presence of
such inflection points in the $FWHM$ graphs at the transition point to
be a generic feature for different trapping curvatures and
dimensionalities, as will be shown below. This feature thus appears to
be a reliable indication for the onset of Mott plateau formation in
confined Bose systems.  Although the $FWHM$ is accessible
experimentally~\cite{esslinger}, limited resolution can make the
location of the inflection point difficult, as we found the $FWHM$
graphs to be rather flat in this region.

\subsection{Parabolic traps in 3D}

We now present results of QMC simulations of the Bose Hubbard model
for bosons confined in 3D parabolic traps. Performing an analysis like
in the 2D case, we find that similar generic features as those
obtained in 2D apply to these 3D systems as well.

\label{sec:sim3D}
\begin{figure}
\begin{center}
\vspace{0.5cm} \includegraphics[width=8cm]{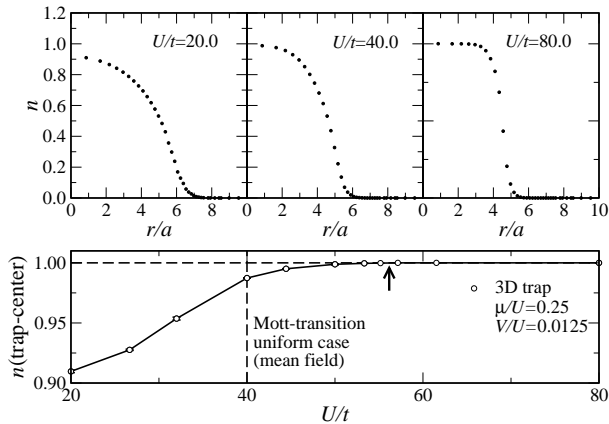}
\caption{ Top: Radial density distribution of bosons within a
three dimensional parabolic trap with curvature $V/U=0.0125$ for
different values of $U/t$ for $\mu/U=0.25$.  Bottom: Evolution of the
density in the trap center while tuning $t$ along a
constant $\mu/U=0.25$ scan.  The arrow indicates the threshold for
Mott plateau formation within the trap, which deviates by about 30\%
from the position of the Mott transition in the uniform case within
mean field theory (vertical dashed line).
}
\label{fig:3Dtrap:density}
\end{center}
\end{figure}

In particular, we consider a parabolic trap with curvature
$V/U=0.0125$, and study the system's states for different values of
$t/U$ along a line of constant $\mu/U=0.25$. For a value of $U/t=20$,
the system is still deep in the superfluid regime, as seen in
Fig.~\ref{fig:3Dtrap:density}, reflecting the increased strength of
the kinetic energy, due to the larger dimensionality.  Upon decreasing
the hopping $t$, a Mott plateau forms in the trap center.  In
Fig.~\ref{fig:3Dtrap:density} we trace the boson density in the
trap center as a function of $U/t$, in order to locate the threshold
for emergence of the Mott plateau region, indicated by the vertical
arrow.  Similar to the 2D case, the central density approaches the
value of 1 with a flat slope.  Within mean field theory~\cite{fisher,
monien} and the local potential approximation, the threshold would be
underestimated by about 30\%, as indicated by the dashed vertical line
in Fig.~\ref{fig:3Dtrap:density}.

\begin{figure} 
\begin{center} 
\vspace{0.5cm} \includegraphics[width=8cm]{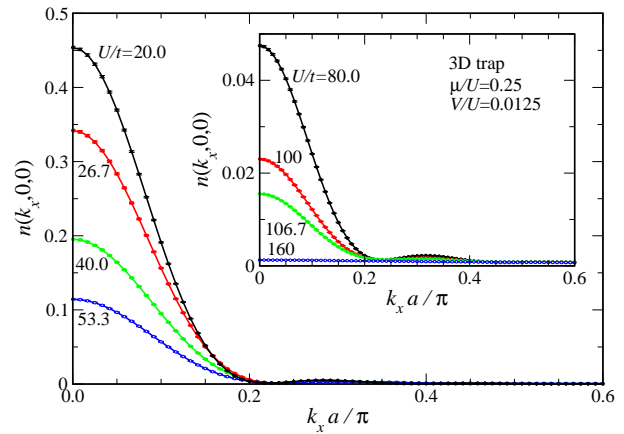}
\caption{ Momentum distribution function of bosons in a
three dimensional parabolic trap with curvature $V/U=0.0125$ along the
line $(0,0,0)$-$(\pi/a,0,0)$ in momentum space, for different values
of $U/t$ while tuning $t$ along the constant $\mu/U=0.25$ scan of
Fig.~\ref{fig:3Dtrap:density}.  Satellite peaks near ${\bf k}=(0.3
\pi/a,0,0)$ appear unrelated to the phase structure within the trap.
} 
\label{fig:3Dtrap:green} 
\end{center} 
\end{figure}

\begin{figure}
\begin{center}
\vspace{0.5cm}
\includegraphics[width=8cm]{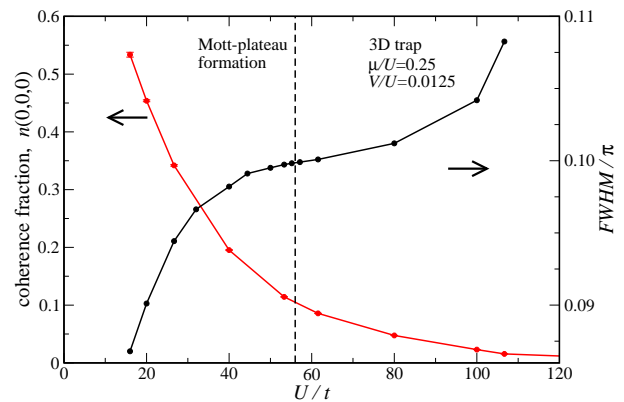}
\caption{ Evolution of the coherence fraction, i.e. the height,
$n(0,0,0)$ and of the full width at half maximum ($FWHM$) of the
coherence peak as a function of $U/t$ while tuning $t$ along the
constant $\mu/U=0.25$ scan of Fig.~\ref{fig:3Dtrap:density} for bosons
in a three dimensional parabolic trap with curvature $V/U=0.0125$. The
threshold for Mott plateau formation is indicated by the dashed line.
}
\label{fig:3Dtrap:fwhm}
\end{center}
\end{figure}

Analyzing the momentum distribution functions shown in
Fig.~\ref{fig:3Dtrap:green}, we observe similar behavior as in the 2D
system: (i) As seen from Fig.~\ref{fig:3Dtrap:fwhm}, at the threshold
for Mott plateau formation, the coherence fraction is still about 10\%
of the overall bosonic density and decreases over a rather broad
range of parameters.

(ii) The $FWHM$ of the coherence peak undergoes a change of curvature
with an inflection point being located at the threshold for
Mott plateau formation.  The presence of this inflection point thus
provides a robust indication of density restructurings also inside 3D
traps.  (iii) As in 2D, we find the presence of satellite peaks to be
unrelated to the local density structure, as seen from
Fig.~\ref{fig:3Dtrap:green}.  (iv) The position of the fine structure
peak is indicative of the spatial extent of the bosonic cloud. In
fact, the broad peak observed at $k_x\approx 0.3 \pi / a$ in
Fig.~\ref{fig:3Dtrap:green} corresponds well to the radial extent
$R\approx 6a$ of the Bose gas.

\subsection{Parabolic traps in 1D}

\begin{figure}
\begin{center}
\vspace{0.5cm}
\includegraphics[width=8cm]{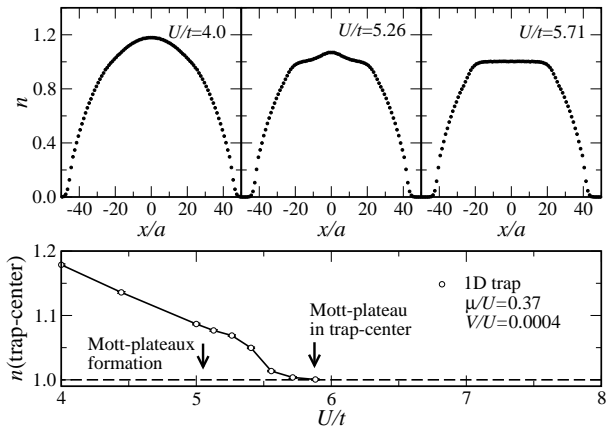}
\caption{ Top: Spatial density distribution of bosons within a
one dimensional parabolic trap with curvature $V/U=0.0004$ for
different values of $U/t$, for $\mu/U=0.37$.  Bottom: Evolution of the
density in the trap center while tuning $t$ along a
constant $\mu/U=0.37$ scan.  The threshold for Mott plateaux formation
within the trap, and merging of the two plateaux are indicated by
vertical arrows.
}
\label{fig:1Dtrap:density}
\end{center}
\end{figure}

Finally, we extend our analysis to the case of a 1D parabolic trap.
Fixing the chemical potential to a value of $\mu/U=0.37$, similar to
the 2D case, we study the system for different values of the hopping
amplitude, $t$.  In the upper part of Fig.~\ref{fig:1Dtrap:density}
the spatial density distribution is shown for three different values
of $U/t$. The evolution of the density in the trap center as a
function of $U/t$ is shown in the lower part of
Fig.~\ref{fig:1Dtrap:density}.

\label{sec:sim1D}

For $U/t=4.0$ the system is in the fully superfluid regime, with no
Mott plateaux present.  Upon increasing $U/t$, there appears a
finite regime, where two Mott plateaux emerge well outside the center
of the trap. These plateaux eventually merge into an extended
Mott plateau at a larger value of $U/t$. The position of these points
is marked by the arrows in the lower part of
Fig.~\ref{fig:1Dtrap:density}. This emergence of an intermediate
regime with two well separated Mott plateaux is expected from the
shape of the first Mott lobe in 1D~\cite{kuehner}, and the chosen
value of $\mu/U=0.37$, and follows using a local potential
approximation, similar to the case of the 2D trap considered
above. The reason why such an intermediate regime is observed in our
1D simulations, but not for the 2D case, is that in 1D the largest
extent of the first Mott lobe has a critical value of
$\mu/U\approx0.10$ which is below our chosen value of $\mu/U=0.37$,
whereas the critical value of $\mu/U=0.42$ in 2D is above that value.

\begin{figure}
\begin{center}
\vspace{0.5cm}
\includegraphics[width=8cm]{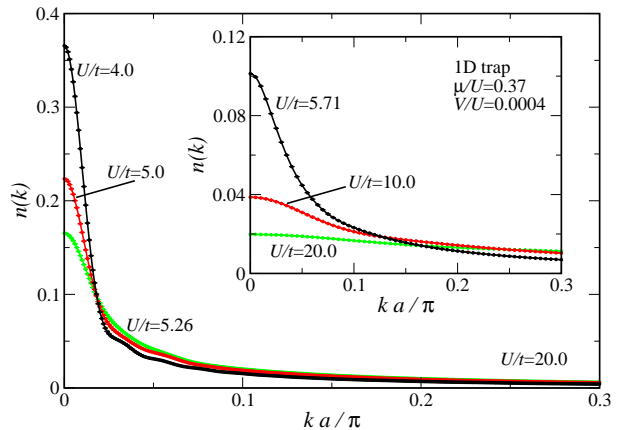}
\caption{ Momentum distribution function of bosons in a
one dimensional parabolic trap with curvature $V/U=0.0004$ along the
line $0$-$\pi/a$ in momentum space, for different values of $U/t$
while tuning $t$ along the constant $\mu/U=0.37$ scan of
Fig.~\ref{fig:1Dtrap:density}.
}
\label{fig:1Dtrap:green}
\end{center}
\end{figure}

The corresponding momentum distribution functions are shown in
Fig.~\ref{fig:1Dtrap:green}. Compared to the higher dimensional cases,
the momentum distribution functions appear broad, indicating larger
incoherent contributions. This is expected, as in 1D long range
coherence cannot develop, even at zero temperature.  Similar to the
higher dimensional cases, we observe broad fine structure peaks in
$n(k)$, restricted however to $U/t$ below the threshold for
Mott plateau formation.  At larger values of $U/t$, such fine
structure is not resolved due to the large incoherent contribution.

\begin{figure}
\begin{center}
\vspace{0.5cm} \includegraphics[width=8cm]{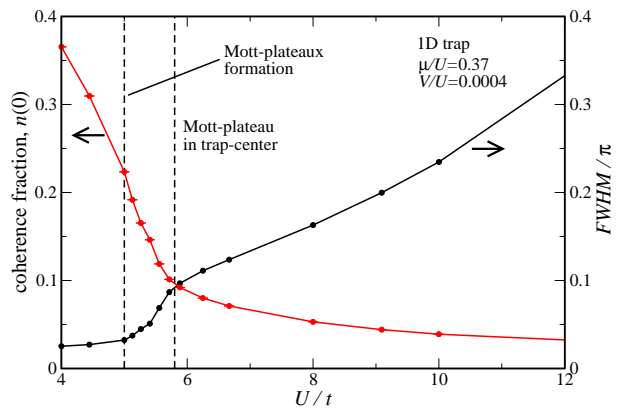}
\caption{ Evolution of the coherence fraction, i.e. the height,
$n(0)$, and of the full width at half maximum ($FWHM$) of the
coherence peak as a function of $U/t$ while tuning $t$ along the
constant $\mu/U=0.37$ scan of Fig.~\ref{fig:1Dtrap:density} for bosons
in a one dimensional parabolic trap with curvature $V/U=0.0004$.  The
threshold for Mott plateaux formation within the trap, and for the
collapse of the two plateaux into a single plateau are indicated by
dashed lines.
}
\label{fig:1Dtrap:fwhm}
\end{center}
\end{figure}

Analyzing the momentum distribution functions as shown in
Fig.~\ref{fig:1Dtrap:fwhm}, we find that the graph of $FWHM$ as a
function of $U/t$ exhibits two characteristic features: The increase
of the slope for $U/t$ near $5.0$ corresponds well to the threshold
for the formation of the two Mott plateaux.  The growth of the two
Mott plateaux regions results in the fast decrease of the coherence
length in this regime.  Beyond $U/t\approx 5.7$, the increase in the
$FWHM$ is reduced, indicating that the two plateaux have merged into a
single plateau, which now grows at only two ends.  The
$U/t$ dependence of the coherence fraction $n(0)$ also indicates the
merging of the two Mott plateaux, by a reduced decrease in $U/t$
beyond $U/t\approx 5.7$.

Recently Kollath {\it et al.}~\cite{kollath} studied the 1D case using
the Density Matrix Renormalization Group method.  Their results are in
perfect agreement with our observations.

\section{Discussion and conclusion}
\label{sec:discussion}
Our quantum Monte Carlo simulations provide insight into the physics of trapped
bosonic systems on lattices. The validity of a local potential
approximation, where local quantities such as the local density or
compressibility depend mainly on the value of the local chemical
potential, is confirmed by an excellent data collapse of local
quantities on single curves.  This single curve is however {\it not}
the same as for the homogeneous bulk system; the differences are
particularly pronounced in the interesting vicinity of the transition layer
between Mott insulators and superfluid in the parabolic trap. 
There the singularities due to
quantum critical behavior are removed and replaced by smooth and broad
features.

While the behavior of the homogeneous system can give a qualitative
overview of the phase structure realized locally inside the parabolic trap,
quantitative results can only be obtained by numerical (quantum Monte Carlo)
simulations of realistic systems.  Results for realistic 2D parabolic traps of a
size comparable to experiments have been presented here, as well as an
analysis of the 1D situation.  Three dimensional simulations have so
far been performed only on systems with linear dimensions 2-3 times
smaller than experimental realizations, but realistic simulations will
be possible in the near future using faster computers and improved
algorithms.

An effective ladder model, which quantitatively models the realistic
trapped system, provides clear evidence for the absence of quantum
critical behavior in parabolic confinement potentials. 
The ladder model allows us to exclude the
randomness imposed on the superfluid ring by the underlying square
lattice structure as the source of this absence of quantum
criticality. Instead, the divergences due to quantum critical
fluctuations are suppressed by the inhomogeneity, and the coupling to the rest of 
the system. It will 
be very interesting to develop an effective action for
this coupling, which might explain the power law behavior observed in
Ref. \cite{muramatsu,bergkvist}.
Furthermore, the observed absence of quantum
criticality in both parabolic traps and ladder models with a gradient
in the chemical potential connecting different Mott plateau regions
calls for future investigations, and extension to fermionic models.

The absence of quantum critical behavior of bosons
in parabolic confinement potentials 
also agrees well with the
fast dynamics of the ``phase transition'' and the observed absence of
``critical slowing down'' of the dynamics in 
experiments~\cite{greiner}. Critical slowing down at a second order
phase transition is caused by the long time scales taken to establish
a uniform order parameter across the whole system. Small ordered
domains, with differently broken $U(1)$ symmetry are quickly formed,
but as these domains grow and merge, the dynamics to establish the
same $U(1)$ symmetry breaking across neighboring domains slows down as
the domains grow in size. The dynamics of the ``quantum phase
transition'' in the trapped system is different: driving the system
from the superfluid to the Mott insulator happens by nucleating a
small Mott domain inside the trap, which then grows as the
depth of the optical lattice is increased. As the Mott phase grows in
volume and the superfluid phase shrinks the ``quantum phase
transition'' is observed. This is however better viewed as a crossover
with changing volume fractions of the two phases than as a phase
transition: the large Mott plateau is always surrounded by a shell of
coherent superfluid. Sweeping back to the superfluid phase by
decreasing the depth of the optical lattice, the Mott insulator melts
and atoms join the superfluid. The dynamics here is not that of two
large domains merging, but that of a single atom joining the coherent
superfluid and there is {\it no critical slowing down} involved in
this process. It might be possible to experimentally observe critical
slowing down in a trap by first driving the system deep into the
Mott insulating region, then kicking it to destroy the phase coherence
in the remaining superfluid shell and afterwards quickly driving it
back into the superfluid. 

Finally, we analyzed the behavior of the
momentum distribution function, which is accessible experimentally
from the interference patterns of absorption images taken after free
expansion of the atomic gas.  We find that the full width at half
maximum of the coherence peak in the momentum distribution function,
due to its relation to the coherence length in the system, yields
valuable information about density restructurings inside the trap. In
particular, we found an inflection point in its graph upon increasing
the lattice depth well at the threshold for central Mott plateau
formation. Since the graph becomes flat in this region, detection of
such features requires high resolution data taken in the crossover
regime.

In contrast,
we found that for flat confinement potentials, realizing
closed box systems, both the full width at half maximum, 
as well as the coherence fraction provide clear signals for 
the Mott transition.
Furthermore, in flat trapping geometries, 
quantum critical fluctuations are
significantly more pronouned, and allow the observation of quantum critical behavior already
on optical lattices of currently available sizes.
We thus expect the possible realization of flat confinement potentials~\cite{zoller_privat} to
significantly ease the detection of true quantum criticality and the interpretation of the
experimental data.

Performing quantum Monte Carlo studies for realistic systems will be important for
interpreting current and future experiments on confined Bose gases in
optical lattices, and for testing our quantitative understanding of 
these systems.
Such understanding will be important,
once analog quantum computers build
from fermionic atoms are available, for which
large scale quantum Monte Carlo simulations will not be possible.
 
\section*{Acknowledgments}

We thank M. Cazalilla, T. Esslinger, A. Ho, A. Muramatsu, 
N. V. Prokof'ev, M. Rigol, and P. Zoller for
fruitful discussions.  Part of the numerical calculations were done
using the SSE application package with generalized directed loop
techniques~\cite{alet} of the ALPS project~\cite{alps}, and performed
on the Asgard Beowulf cluster at ETH Z\"urich.  S.W. and
M.T. acknowledge hospitality of the Kavli Institute for Theoretical
Physics, Santa Barbara, and M.T. also the Aspen Center for Physics.
This research was supported in part by the National Science Foundation
under Grant No. PHY99-07949.  S.W., F.A. and M.T. acknowledge support
from the Swiss National Science Foundation.  G.B. is supported by the
NSF-CNRS cooperative grant \#12929 and thanks the NTNU, the Complex
Group and Norsk Hydro for their hospitality and generosity during a
sabbatical stay.

\end{document}